\documentclass{PoS}
\usepackage{amssymb,amsmath}
\title{Excitation of wakefield around pulsars}

\ShortTitle{Wakefield generation}

\author{Vazha I. Berezhiani\\
        School of Physics, Free University of Tbilisi, 0183-Tbilisi, Georgian\\
    }

\author{\speaker{Zaza Osmanov}\\
        School of Physics, Free University of Tbilisi, 0183-Tbilisi, Georgia\\
        E-mail: \email{z.osmanov@freeuni.edu.ge}}

\author{Milivoj Belic\\
        Science Program, Texas A\&M University at Qatar, PO Box 23874 Doha
        }

        \abstract{We study the generation of the wakefields by means
          of the high energy radiation of pulsars. The problem is
          considered in the framework of a one dimensional
          approach. We linearize the set of governing equations
          consisting of the momentum equation, continuity equation and
          Poisson equation and show that a wavelike structure will
          inevitably arise relatively close to the pulsar.}

\FullConference{The Modern Physics of Compact Stars 2015\\
        30 September 2015 - 3 October 2015\\
        Yerevan, Armenia}

\begin{document}

\section{Introduction}
It is strongly believed that media around pulsars are efficiently
perturbed by their powerful high energy emission, which in turn is
directly related to the neutron star's rotation. The total energy of
these objects can provide is $W_{\rm tot}\ = I\Omega^2/2$, where
$I=2MR_{\star}^2/5$ is pulsar's moment of inertia,
$M\approx 1.5\times M_{\odot}$ and $R_{\star}\approx 10^6$cm are
pulsar's mass and radius respectively,
$M_{\odot}\approx 2\times 10^{33}$g is the solar mass, $\Omega=2\pi/P$
is the angular velocity of rotation and $P$ is the corresponding
period. It is evident that the rotational energy budget of the pulsar
increases for rapidly rotating objects and for millisecond pulsars
this might become very large. Indeed, by applying the typical parameters of
millisecond pulsars, one can see that this energy is of the order of
\begin{equation}
\label{etot} W_{\rm tot}\approx 2.4\times 10^{52}
\times\left(\frac{P}{0.01s}\right)^{-2}\times\left(\frac{M}{1.5M_{\odot}}\right)
~{\rm ergs}.
\end{equation}
As we see, if even a tiny fraction of $W_{\rm tot}$ is transformed
into radiation, its role might be significant.

It is well known from observations that all pulsars slows down, which
in turn means that their released rotational energy can strongly
influence their overall emission pattern. The corresponding parameter
is called the slow down rate $\dot{P}\equiv
dP/dt<0$. It is straightforward to estimate the energy released in
unit of time (slow-down luminosity) $L_{\rm sd}\equiv\dot{W_{\rm tot}} =
I\Omega|\dot{\Omega}|$, where  $\dot{\Omega}\equiv
d\Omega/dt=-2\pi\dot{P}/P^2$
\begin{equation}
\label{lsd} L_{\rm sd}\approx 9.5\times 10^{38}
\times\left(\frac{P}{0.01{\rm s}}\right)^{-3}\times
\left(\frac{\dot{P}}{10^{-13}{\rm ss}^{-1}}
\right)\times\left(\frac{M}{1.5M_{\odot}}\right) ~{\rm ergs}\, \rm{s}^{-1}.
\end{equation}
As we see, even if a tiny fraction of this huge amount of power
transforms into that of emission one might have interesting
consequences. It is commonly assumed that on average only $1\%$ of rotational
energy is converted to radiation, therefore its luminosity is given by
\begin{equation}
\label{lb} L\approx 9.5\times
10^{36}\times\left(\frac{\kappa}{0.01}\right)
\times\left(\frac{P}{0.01{\rm s}}\right)^{-3}\times
\left(\frac{\dot{P}}{10^{-13}{\rm ss}^{-1}}
\right)\times\left(\frac{M}{1.5M_{\odot}}\right)
~{\rm ergs}\, \rm{s}^{-1}.
\end{equation}
where we have taken into account the expression for the bolometric
luminosity $ L\approx\kappa L_{\rm sd}$, where $\kappa$ is the fraction of the
rotational energy that is converted into radiation.
Such a huge radiation power will inevitably undergo the overall
pattern around a pulsar and since the radiated energy is very high
such a pattern probably might be seen also on distant regions from the
central object - thus in the nebular structure.

Apart from the astrophysical example of pulsars discussed above, it is
clear that similar physics is encountered in the interactions of
strong electromagnetic pulses with plasma in laboratory.  In
particular, Tajima \& Dawson \cite{tajima} considered short laser
radiation and its effect on plasmas. The authors examined radiation
with power density of the order of $10^{18}$W/cm$^2$ irradiating
a plasma with particle number densities of the order of $10^{18}$cm$^{-3}$
\cite{tajima}. The group velocity of the electromagnetic radiation is
given by
\begin{equation}
\label{vg}
\upsilon_g^{EM} =
c\left(1-\frac{\omega_p^2}{\omega^2}\right)^{1/2}<c,
\end{equation}
where $\omega_p\equiv\sqrt{4\pi n_0e^2/m}$ is the plasma frequency,
$n_0$ is the number density of plasma particles, $e$ and $m$ are
respectively the charge and mass of an electron and $\omega$ is the
photon frequency.  According to Ref. \cite{tajima}  the interaction of
the radiation with the group velocity \eqref{vg} with plasma creates a
 wakefield with  the phase velocity
\begin{equation}
\label{vp} \upsilon_p = \frac{\omega_p}{k_p} = \upsilon_g^{EM},
\end{equation}
where $k_p$ is the wave-number of the plasmon. From Eq.~\eqref{vp} it is
evident that the Lorentz factor of the plasmon is of the order of
$\gamma \approx \omega/\omega_p$, which, as we will see later, could
be very large in the
realistic astrophysical scenarios. As
explained in Ref.~\cite{tajima} the electromagnetic waves
lead to transverse oscillations of electrons. In the nonrelativistic
limit the corresponding average energy of oscillations might be
estimated as to be
\begin{equation}
\label{dw} \langle\Delta W_T\rangle\approx
\frac{e^2}{2m\omega^2}\langle E_y^2\rangle,
\end{equation}
where $E_y$ is the transverse component of the electric field. The
momentum the electrons pick up is of the order of \cite{tajima}
\begin{equation}
\label{dp} \langle\Delta p_x\rangle\approx \frac{\langle\Delta
W_T\rangle}{c}.
\end{equation}
During the time duration of the pulse, $\tau$, the electrons are
displaced by the lengthscale $\Delta x = \langle\Delta
\upsilon_x\tau\rangle$. After passing the pulse the charge
separation pulls the electrons back and as a result the plasma
oscillations are generated. Due to propagation of the excited
wakefield the electrons are trapped and accelerated. In the rest
frame of plasmons the electrons gain energy, $W\approx e\varphi$,
which in the lab frame transforms to \cite{tajima}
\begin{equation}
\label{Wmax} W_{max}\approx2\gamma^2\varphi,
\end{equation}
where $\varphi$ is the potential in the wake. From Eq. (\ref{Wmax}) it
is evident that the wakefield might efficiently accelerate particles.
The authors of Ref.~\cite{tajima} have shown that short duration
pulses of electromagnetic radiation might efficiently induce the
Langmuir waves.

On the other hand, unlike the laboratory processes, in
astrophysical situations radiation of astrophysical sources is
characterized by a broad band emission. Despite this difference the
major mechanism of wakefield generation still works, because during
the interaction of very high energy radiation with plasma particles
the ions and lighter electrons behave differently.  This difference
will inevitably lead to charge separation creating an electrostatic
field, which in turn will act on the particles. The interaction of
high intensity (of the order of $10^{27}$ergs cm$^{-2}$s$^{-1}$)
radiation with plasmas having density $n$ has been studied by Zampieri
et al. \cite{zamp}. To study the interaction of high energy radiation
with the plasma particles the authors considered the following
equations
\begin{eqnarray}
\label{zamp1}
\gamma_e\frac{d}{dt}\left(\gamma_em_e\upsilon_e\right)
&=& -\gamma_eZeE+F_{\rm rad},\\
\label{zamp2} \gamma_i\frac{d}{dt}\left(\gamma_im_i\upsilon_i\right)
&=& \gamma_iZeE,
\end{eqnarray}
where by $\gamma_e$ and $m_e$ denote the Lorentz factor and mass of
electrons and $\gamma_i$ and $m_i$ are the same quantities for ions,
$\upsilon_{e,i}$ are their velocities, $Z$ is the atomic
number of ions, $E$ is the induced electric field and
\begin{equation}
\label{compt2} F_{\rm rad} =
-\int\frac{\epsilon_1\mu_1-\epsilon\mu}{c}\gamma_e\left(1-\frac{\upsilon_e}{c}\mu\right)\frac{I_{\epsilon}}
{\epsilon}\frac{d\sigma}{d\Omega_s}d\epsilon d\Omega d\Omega_s,
\end{equation}
is the radiative force acting on electrons, $I_{\epsilon}$ is the
specific intensity of radiation,  $\Omega$ denotes the solid angle
subtended by the astrophysical source and $\Omega_s$ is the solid
angle over the scattering angle, $\epsilon$ and $\epsilon_1$ are the
energies of photons in the laboratory frame (LF) of reference, $\mu$
($\mu_1$) are cosines of angles between radial and incident
directions respectively and $d\sigma/d\Omega$ is the
Klein-Nishina differential cross section \cite{blgl}. By applying
the Gauss's law, $E = 4\pi Zen\Delta x$, where $n$ is the number
density of charges and $\Delta x = r_e-r_i$ is the charge separation
the authors reduce the system of equations \eqref{zamp1}-\eqref{compt2} to
\begin{equation}
\label{zamp3} \frac{d^2x_e}{dt^2} = -\frac{4\pi Z^2e^2n}{m_e}\Delta
x+\frac{\sigma_TF}{m_ec},
\end{equation}
\begin{equation}
\label{zamp4} \frac{d^2x_i}{dt^2} = \frac{4\pi Z^2e^2n}{m_i}\Delta
x,
\end{equation}
where $F\equiv L/4\pi r_e^2$ is the radiation intensity and
$\sigma_T\approx 6.65\times 10^{-25}$cm$^2$ is the Thomson cross
section. By solving and analyzing these equations the authors have
argued that radiation-induced electric field might efficiently
influence the maximum energy of plasma particles.

The generation of wakefield has been considered in a series of
works. Gorbunov \& Kirsanov \cite{gorb} have examined the nonlinear
excitation of longitudinal electrostatic waves in plasmas. These
authors studied a set of equations governing the process of wakefield
excitation by applying  the hydrodynamic equation
\begin{equation}
\label{gorb1} \frac{\partial p_{\perp}}{\partial
t}+\upsilon\frac{\partial p_{\perp}}{\partial x} =
eE-\frac{e}{c}\upsilon_{\parallel}B
\end{equation}
and two Maxwell's equations
\begin{eqnarray}
\label{gorb2} \frac{\partial E}{\partial x} &=&
-\frac{1}{c}\frac{\partial B}{\partial t},\\
\label{gorb3} \frac{\partial B}{\partial x} &=&
-\frac{1}{c}\frac{\partial E}{\partial t}+\frac{4\pi
e}{c}n\upsilon_{\perp},
\end{eqnarray}
where $\upsilon_{\perp}$ and $\upsilon_{\parallel}$ are the
transverse and longitudinal velocity components of electrons and
$p_{\perp}$ is the corresponding transverse component of momentum.
Assuming that the transverse velocity is given by
\begin{equation}
\label{gorb4} \upsilon_{\perp}(x,t) =
\frac{1}{2}\left[a(\chi)e^{-i\omega t+ikx}+a^{\ast}(\chi)e^{i\omega
t-ikx}\right],
\end{equation}
where the dispersion relation is $\omega^2=k^2c^2+\omega_p^2$ and  $\chi
= x-\upsilon_gt$, where $\upsilon_g$ is the group velocity of the
wave, the authors find from
Eqs. (\ref{gorb1})-(\ref{gorb4})  the following low frequency
perturbation
\begin{equation}
\label{gorb5} \frac{\delta n_0}{n_0} =
\frac{1}{4\upsilon_g^2}\left(1-2\sin^2\left[k_p\left(\frac{\chi}{2}-\frac{L}{4}\right)\right]\right),
\end{equation}
for $L/2>\chi>-L/2$ and
\begin{equation}
\label{gorb6} \frac{\delta n_0}{n_0} =
\frac{1}{2\upsilon_g^2}\sin\left(k_p\chi\right)\sin\left(\frac{k_pL}{2}\right),
\end{equation}
for $\chi<-L/2$, where $k_p = \omega_p/\upsilon_g$. On the basis of
these results Gorbunov \& Kirsanov \cite{gorb} showed that the
wakefield excited behind the packet can strongly affect particle
acceleration. They argued that particles injected into the plasma are
accelerated quite efficiently. The change in the sign of electric
field leads to a deceleration phase. The overall effect is that the
electrons trapped in the wakefield region have maximal energy given by
\begin{equation}
\label{gorb7} \epsilon_{max} =
\epsilon_0\gamma\frac{eE}{m_e\omega_pc}\left(1+\left[1+\frac{2m_ek_p}{eE\gamma}\right]^{1/2}\right),
\end{equation}
where $\epsilon_0$ is the initial energy of electrons, $e$ is the
elementary charge.

In Ref. \cite{bulanov} the authors have examined the case of
relativistic intense pulse. To study the role of
linearly polarized electromagnetic waves on the ambient plasma these
authors considered the equation governing the evolution of a
dimensionless vector potential, $A\equiv p_{\perp}/mc$, describing
the pulse $A=0.5[a(\chi, t)\exp\left(-e\omega_0t+ik_0x\right)+c.c.]$,
where $k_0$ is the wave vector of radiation and $\omega_0^2 =
k_o^2c^2-\omega_p^2$, $\chi = x-\upsilon_g t$,
$\upsilon_g=c^2k_0/\omega_0$.
This equation, given by Ref. \cite{bulanov}, reads
\begin{eqnarray}
\label{bul1} 2i\omega_0\frac{\partial a}{\partial
t}+\frac{\omega_p^2}{\omega_0^2}c^2\frac{\partial^2 a}{\partial
\chi^2}+2\upsilon_g\frac{\partial a}{\partial t\partial\chi} &=&
-\omega_p^2\frac{\psi_0}{1+\psi_0}a,\\
\label{bul2} \frac{d^2\psi_0}{d\chi^2}-
k_p^2\frac{1+|a|^2/2-\left(1+\psi_0\right)^2}{2\left(1+\psi_0\right)^2}
&=& 0,
\end{eqnarray}
where, as above, $k_p = \omega_p/\upsilon_g$ and $\psi_0$ is the low
frequency component of the potential.  A laser pulse with a wavelength
of the order of $10\mu$m and intensity $10^{17}$W/cm$^2$ having an
edge length of $0.1$ps was considered. Using Eqs.~\eqref{bul1} and
\eqref{bul2} it was shown that the pulse will accelerate the electrons
up to eneregies $1$~GeV over a distance $50$~cm in a plasma medium with
number density of the order of $10^{15}$cm$^{-3}$.

The problem of generation of wakefield has been examined by
Berezhiani \& Murusidze in Ref.~\cite{bm90} . To address this problem
the authors used the Maxwell's equations in combination with the
equations of motion and derived the following equation for
  electron momentum
\begin{equation}
\label{berezh1} \frac{\partial^2 p}{\partial t^2} -\Delta
p+\nabla\left(\nabla\cdot p\right)+\frac{\partial}{\partial
t}\nabla\sqrt{1+p^2}+\frac{p}{\sqrt{1+p^2}}\left[1+\partial
\left(\nabla\cdot p\right)/\partial t+\Delta\sqrt{1+p^2}\right]=0,
\end{equation}
which for the transparent plasma medium reduces to
\begin{equation}
\label{berezh2}
\frac{d^2y}{dx^2}=\frac{1}{2}\left(\frac{\gamma_{\perp}^2}{y^2}-1\right),
\end{equation}
where $\gamma_{\perp}^2 = 1+p_{\perp}$, $x = k_p\left(z-ct\right)$,
$y = \sqrt{\gamma_{\perp}^2+p_{\parallel}^2}-p_{\parallel} = 1+\Phi$
and $\Phi\approx e\Phi/mc^2$ is the dimensionless potential. By
analyzing Eq.~\eqref{berezh2} it was shown that the
maximum increase of energy of trapped electrons is given by
\begin{equation}
\label{berezh3} \Delta E=2\gamma_g^2mc^2\Delta\Phi_{max},
\end{equation}
where $\gamma_g = \left(1-\upsilon_g^2/c^2\right)^{-1/2}$.
Therefore, Berezhiani \& Murusidze~\cite{bm90}  have shown that longitudinal
oscillations with relativistic phase velocities lead to the excitation of the wakefield.

Finally, we note that Sprangle et al. \cite{sprangle} developed a
nonlinear theory of laser-plasma interactions and studied the
relativistic optical guiding, nonlinear excitation of wakefields and
generation of coherent harmonic radiation.

In this paper we consider the possibility of excitation of wakefields
via the interaction of the pulsar high energy emission with plasma
surrounding the pulsar.  Generally speaking,  pulsars are emitting
in narrow channels, which in turn are rotating. Therefore any
irradiated area surrounding the pulsar experiences the radiation
pressure periodically, which can lead to generation of wakefields.

The structure of the paper is as follows. In Sec. II we develop an
analytical method for studying the generation of wakefields in
media, periodically irradiated by the pulsar's high-energy emission.
In Sec. III we present our results and summarize them in the Sec. IV.

\section{Theoretical background}
In this section we consider the set of equations which govern the
generation of wakefields. In the framework of our model this structure
is formed due to the high-energy radiation pressure of a pulsar. On
the other hand this pressure is created by means of the Compton
interaction of radiation photons and electrons in the medium
surrounding the neutron star. The corresponding energy of scattered
electrons has the following form
(see Ref. \cite{lightman})
\begin{equation}
\label{compt1} \epsilon_1' =
\frac{\epsilon'}{1+\left(\epsilon'/mc^2\right)\left(1-\cos\chi'\right)},
\end{equation}
where $\epsilon'$ and $\epsilon_1'$ are the photon energies before
and after scattering respectively in the rest frame of the electron,
$m$ is its mass, $c$ is the speed of light and $\chi'$ is the angle
between the incident and scattered photon directions.

In the framework of the Thomson scattering,
$\epsilon_1'\approx\epsilon'$, Eq. (\ref{compt2}) reduces to~\cite{zamp}
\begin{equation}
\label{comp3} F_{_{rad}} = \sigma_T\frac{F}{c},
\end{equation}
where $F$ is given by
\begin{equation}
\label{comp4} F = \int I_{\epsilon}\mu d\epsilon d\Omega =
\frac{L}{4\pi r^2\left(1-\cos\left[\theta/2\right]\right)},
\end{equation}
where it has been taken into account that each radiation channel
carries half of the luminosity. The corresponding  solid angle of
the cone is $2\pi\left(1-\cos\left[\theta/2\right]\right)$ and
$\theta$ is the opening angle of the radiation cone.

The system of equations governing the generation of wakefields
consists of the momentum equation \cite{gorb}
\begin{equation}
\label{eul1} \frac{\partial\upsilon_{_{\parallel}}}{\partial
t}+\upsilon_{_{\parallel}}\frac{\partial\upsilon_{_{\parallel}}}{\partial
x} = -\frac{e}{m}\frac{\partial\varphi}{\partial x}+f_{\rm rad},
\end{equation}
the continuity equation
\begin{equation}
\label{con1} \frac{\partial n}{\partial t}+\frac{\partial}{\partial
x}\left(n\upsilon_{_{\parallel}}\right) = 0,
\end{equation}
and the Poisson equation
\begin{equation}
\label{poi1} \frac{\partial^2\varphi}{\partial x^2} = -4\pi
e\left(n-n_0\right),
\end{equation}
where $\upsilon_{_{\parallel}}$ is the velocity component parallel to
the $x$-axis, $\varphi$ denotes the charge separation potential,
$f_{\rm rad}=F_{\rm rad}/m$, $n$ is the electron's number density and
$n_0$ is the unperturbed number density.

We solve the problem of generation of the wakefield
by linearizing the governing equations, which  might be
easily reduced to
\begin{equation}
\label{eq1} \frac{\partial^2\delta n}{\partial t^2} +
\omega_p^2\delta n = -n_0\frac{\partial f_{\rm rad}}{\partial x},
\end{equation}
where by $\delta n$ we denote the density perturbation.

Since the excited modes are traveling waves it is convenient to
introduce the variable $\xi\equiv x-ct$, which leads to the
following form of the poisson equation
\begin{equation}
\label{poi2} \frac{\partial^2\varphi}{\partial\xi^2} = -4\pi e\delta
n.
\end{equation}
By applying this expression to Eq. (\ref{eq1}) one gets
\begin{equation}
\label{eq2} \frac{\partial^4 \varphi}{\partial\xi^4} +
\frac{\omega_p^2}{c^2}\frac{\partial^2\varphi}{\partial\xi^2} =
\frac{m\omega_p^2}{ec^2}\frac{\partial f_{\rm rad}}{\partial\xi},
\end{equation}
which straightforwardly can be reduced to
\begin{equation}
\label{eq3} \frac{\partial^2\varphi}{\partial\xi^2} +
\frac{\omega_p^2}{c^2}\varphi = \frac{m\omega_p^2}{ec^2}\int
f_{\rm rad}(\xi)d\xi.
\end{equation}
By assuming that $f_{\rm rad}(\xi)$ is a step-like function, the
solution of Eq.~\eqref{eq3} can be expressed as follows
\begin{equation}
\label{poten} \varphi(\beta) =
\frac{mf_{{\rm rad},0}\xi_0}{e}\left[1-\frac{\sin\alpha}{\alpha}\cos\left(\alpha\{\beta-1\}\right)
-\frac{1-\cos\alpha}{\alpha}\sin\left(\alpha\{\beta-1\}\right)\right],
\end{equation}
where $f_{{\rm rad},0}$ is the radiation reaction force, $\xi_0 =
c\theta/\Omega$, $\alpha = \omega_p\xi_0/c$ and $\beta = \xi/\xi_0$.

\section{Discussion}


We now examine generation of wakefields inside the nebula around a
pulsar. For this purpose we consider the Crab nebula parameters:
mass $M_{_{Nb}}\sim 4M_{\odot}$ (see Ref. \cite{crabneb}) and radius
$R_{_{Nb}}\sim 1.7$pc, which thus define the average number density
 of matter $n_{_{Nb}}\sim 3M_{_{Nb}}/4\pi R_{_{Nb}}^3\sim8$ cm$^{-3}$.
We will assume below that escaping particles do
not change the average density of the Crab nebula. The
Goldreich-Julian number density is given by \cite{gj}
\begin{equation}
\label{GJ} n_{_{GJ}}=\frac{\Omega B_{st}}{2\pi
ec}\times\left(\frac{R_{\star}}{r}\right)^3,
\end{equation}
where $B_{\star}\approx 1.8\times 10^{12}\times\sqrt{P\dot{P}}$G is
the surface magnetic field of the pulsar. On the other hand, outside the
rotationally driven region (magnetosphere) the behaviour of magnetic
field is $1/r$ (see Ref. \cite{manchester}), which leads to the
following condition
\begin{equation}
\label{rcr} r\gg 1.2\times
10^{-6}\times\left(\frac{P}{0.01s}\right)^{-5/2}\times\left(\frac{\dot{P}}{10^{-13}ss^{-1}}
\right)^{1/2}\times\left(\frac{R_{\star}}{10^6cm}\right)^3\times\left(\frac{n_{_{Nb}}}{10cm^{-3}}\right)^{-1}
{\rm AU}
\end{equation}

It is clear that the effect we are interested in strongly depends on
the luminosity of a pulsar. In general there is a special class of
pulsars, the so called $X$-ray pulsars, which exhibit extremely high
values of luminosity. It is believed that their emission has a
synchrotron origin. This process is maintained due to the quasi-linear
diffusion (see Ref. \cite{mo}), when the opening angle of a radiation
cone is expressed as \cite{machus}
\begin{equation}
\label{angle} \theta\approx
0.14\times\left(\frac{0.01s}{P}\right)^{1/2}\times\left(\frac{R_{\star}}{10^6cm}\right)^{1/2}
 {\rm rad}.
\end{equation}
If the electrons are trapped inside the wakefield, they will
inevitably experience the potential difference [see Eqs.
\eqref{comp3},\eqref{comp4},\eqref{lb} and \eqref{angle}]
\begin{eqnarray}
\varepsilon({\rm eV})&\approx& 5.4\times
\left(\frac{\kappa}{0.01}\right)
\times\left(\frac{P}{0.01{\rm s}}\right)^{-5/2}
\times\left(\frac{\dot{P}}{10^{-13}{\rm ss}^{-1}}\right)
\nonumber\\
&\times&\left(\frac{M}{1.5M_{\odot}}\right)
\times\left(\frac{R_{\star}}{10^6{\rm cm}}\right)^{-1/2}
\times\left(\frac{r}{0.01 {\rm AU}}\right)^{-2}.
\end{eqnarray}
It is seen that, for example,
on distances $0.01$ AU from the Crab pulsar the
potential difference will be of the order of $4$ eV;
for $r\sim 0.001$ AU we find  $\varepsilon({\rm eV})\sim 0.4$ keV.

Since the most energetic pulsars are the so called newly born young
millisecond pulsars \cite{carroll}, it is interesting to
estimate the wakefield generation for this particular class of
pulsars too. By considering $P\sim 0.001$s and $\dot{P}\sim
10^{-12}$ss$^{-1}$ one can show that for $0.01$ AU the potential
difference will be $\sim2$ keV.

\section{Summary} \label{sec:summary}

\begin{enumerate}
\item We have examined the excitation of wakefield in the media
  surrounding a pulsar. The problem has been studied in a one
  dimensional approximation, assuming that the particles follow along
  the magnetic field lines. Therefore,  we started from the
  one-dimensional form of the  momentum equation,
  the continuity equation and the Poisson equation, linearized them,
  and solved the resulting equations in the case where the radiation
  force has a simple step-like form.

      \item The model has been applied to the Crab-like pulsars and
      the newly born young millisecond pulsars. It was shown that in
      case of the Crab-like pulsars, trapped electrons will be subject
      to the potential difference of the order of
      $0.4$ keV, whereas for the newly born pulsars this value can be
      even higher, of the order $2$ keV.

 \end{enumerate}

\section*{Acknowledgments}

VIB and MB were partially supported by the NPRP 6-021-1-005 project
from the Qatar National Research Fund (a member of the Qatar
Foundation). ZO was partially supported by the Shota Rustaveli
National Science Foundation grant (N31/49).

\end{document}